\documentclass[times, twocolumn]{aastex631_arxiv}
\usepackage{CJK}
\usepackage{graphicx}
\usepackage{enumerate}
\usepackage{amssymb, amsmath}
\usepackage{natbib}
\usepackage{color}
\usepackage{ulem}
\usepackage{dblfnote}
\usepackage{appendix}
\usepackage{hyperref}
\usepackage[stable]{footmisc}
\usepackage{comment}
\usepackage{footnote}
\usepackage{realboxes}

\usepackage[nodayofweek]{datetime}
\newdateformat{monthyearday}{%
  \THEYEAR\ \monthname[\THEMONTH] \twodigit{\THEDAY}}
  
\submitjournal{AAS Journals}
\shorttitle{Keck/NIRC2 follow-up of Subaru/IRD-SSP RV survey}
\shortauthors{Uyama et al.}

\begin{document}

\title{Direct Imaging Explorations for Companions around Mid-Late M Stars from the Subaru/IRD Strategic Program}

\correspondingauthor{Taichi Uyama}
\email{tuyama@ipac.caltech.edu}

\author[0000-0002-6879-3030]{Taichi Uyama}
    \affiliation{Infrared Processing and Analysis Center, California Institute of Technology, 1200 E. California Blvd., Pasadena, CA 91125, USA}
    \affiliation{NASA Exoplanet Science Institute, Pasadena, CA 91125, USA}
    \affiliation{National Astronomical Observatory of Japan, 2-21-1 Osawa, Mitaka, Tokyo 181-8588, Japan}

\author[0000-0002-5627-5471]{Charles Beichman}
    \affiliation{NASA Exoplanet Science Institute, Pasadena, CA 91125, USA}
    \affiliation{Infrared Processing and Analysis Center, California Institute of Technology, 1200 E. California Blvd., Pasadena, CA 91125, USA}
\author[0000-0002-4677-9182]{Masayuki Kuzuhara}
    \affiliation{Astrobiology Center, 2-21-1 Osawa, Mitaka, Tokyo 181-8588, Japan}
    \affiliation{National Astronomical Observatory of Japan, 2-21-1 Osawa, Mitaka, Tokyo 181-8588, Japan}
\author[0000-0001-8345-593X]{Markus Janson}
    \affiliation{Department of Astronomy, Stockholm University, AlbaNova University Center, SE-10691, Stockholm, Sweden}

\author[0000-0001-6181-3142]{Takayuki Kotani}
    \affiliation{Astrobiology Center, 2-21-1 Osawa, Mitaka, Tokyo 181-8588, Japan}
    \affiliation{National Astronomical Observatory of Japan, 2-21-1 Osawa, Mitaka, Tokyo 181-8588, Japan}
    \affiliation{Department of Astronomical Science, The Graduate University for Advanced Studies, SOKENDAI, 2-21-1Osawa, Mitaka, Tokyo 181-8588, Japan}
\author[0000-0002-8895-4735]{Dimitri Mawet}
    \affiliation{Department of Astronomy, California Institute of Technology, 1200 E. California Blvd., Pasadena, CA 91125, USA}
    \affiliation{Jet Propulsion Laboratory, California Institute of Technology, 4800 Oak Grove Dr., Pasadena, CA, 91109, USA}
\author[0000-0001-8033-5633]{Bun'ei Sato}
    \affiliation{Department of Earth and Planetary Sciences, Tokyo Institute of Technology, Meguro-ku, Tokyo 152-8551, Japan}
\author[0000-0002-6510-0681]{Motohide Tamura}
    \affiliation{Department of Astronomy, The University of Tokyo, 7-3-1, Hongo, Bunkyo-ku, Tokyo 113-0033, Japan}
    \affiliation{Astrobiology Center, 2-21-1 Osawa, Mitaka, Tokyo 181-8588, Japan}
    \affiliation{National Astronomical Observatory of Japan, 2-21-1 Osawa, Mitaka, Tokyo 181-8588, Japan}
\author[0000-0001-6309-4380]{Hiroyuki Tako Ishikawa}
    \affiliation{Astrobiology Center, 2-21-1 Osawa, Mitaka, Tokyo 181-8588, Japan}
    \affiliation{National Astronomical Observatory of Japan, 2-21-1 Osawa, Mitaka, Tokyo 181-8588, Japan}

\author[0000-0001-6279-0595]{Bryson Cale}
    \affiliation{Jet Propulsion Laboratory, California Institute of Technology, 4800 Oak Grove Dr., Pasadena, CA, 91109, USA}
    \affiliation{Infrared Processing and Analysis Center, California Institute of Technology, 1200 E. California Blvd., Pasadena, CA 91125, USA}
\author[0000-0002-7405-3119]{Thayne Currie}
    \affiliation{Subaru Telescope, 650 N. Aohoku Place, Hilo, HI 96720, USA}
    \affiliation{Department of Physics and Astronomy, University of Texas-San Antonio, San Antonio, TX, USA}
\author[0000-0002-7972-0216]{Hiroki Harakawa}
    \affiliation{Subaru Telescope, 650 N. Aohoku Place, Hilo, HI 96720, USA}
\author[0000-0002-1493-300X]{Thomas Henning}
    \affiliation{Max-Planck-Institut f\"{u}r Astronomie, K\"{o}nigstuhl 17, 69117 Heidelberg, Germany}
\author[0000-0003-3618-7535]{Teruyuki Hirano}
    \affiliation{Astrobiology Center, 2-21-1 Osawa, Mitaka, Tokyo 181-8588, Japan}
\author[0000-0003-0786-2140]{Klaus Hodapp}
    \affiliation{University of Hawaii, Institute for Astronomy, 640 N. Aohoku Place, Hilo, HI 96720, USA}
\author[0000-0003-4676-0251]{Yasunori Hori}
    \affiliation{Astrobiology Center, 2-21-1 Osawa, Mitaka, Tokyo 181-8588, Japan}
\author[0000-0003-1906-4525]{Masato Ishizuka}
    \affiliation{Department of Astronomy, The University of Tokyo, 7-3-1, Hongo, Bunkyo-ku, Tokyo 113-0033, Japan}
\author{Shane Jacobson}
    \affiliation{University of Hawaii, Institute for Astronomy, 640 N. Aohoku Place, Hilo, HI 96720, USA}
\author[0000-0002-8607-358X]{Yui Kasagi}
    \affiliation{Department of Astronomical Science, The Graduate University for Advanced Studies, SOKENDAI, 2-21-1Osawa, Mitaka, Tokyo 181-8588, Japan}
    \affiliation{National Astronomical Observatory of Japan, 2-21-1 Osawa, Mitaka, Tokyo 181-8588, Japan}
\author[0000-0002-5486-7828]{Eiichiro Kokubo}
    \affiliation{National Astronomical Observatory of Japan, 2-21-1 Osawa, Mitaka, Tokyo 181-8588, Japan}
\author{Mihoko Konishi}
    \affiliation{University of Hawaii, Institute for Astronomy, 640 N. Aohoku Place, Hilo, HI 96720, USA}
\author[0000-0002-9294-1793]{Tomoyuki Kudo}
    \affiliation{Subaru Telescope, 650 N. Aohoku Place, Hilo, HI 96720, USA}
\author{Takashi Kurokawa}
    \affiliation{Astrobiology Center, 2-21-1 Osawa, Mitaka, Tokyo 181-8588, Japan}
    \affiliation{Institute of Engineering, Tokyo University of Agriculture and Technology, 2-24-16 Nakacho, Koganei, Tokyo 184-8588, Japan}
\author[0000-0001-9194-1268]{Nobuhiko Kusakabe}
    \affiliation{Astrobiology Center, 2-21-1 Osawa, Mitaka, Tokyo 181-8588, Japan}
    \affiliation{National Astronomical Observatory of Japan, 2-21-1 Osawa, Mitaka, Tokyo 181-8588, Japan}
\author[0000-0003-2815-7774]{Jungmi Kwon}
    \affiliation{Department of Astronomy, The University of Tokyo, 7-3-1, Hongo, Bunkyo-ku, Tokyo 113-0033, Japan}
\author{Masahiro Machida}
    \affiliation{Department of Earth and Planetary Sciences, Faculty of Sciences, Kyushu University, Fukuoka, Fukuoka 819-0395, Japan}
\author[0000-0002-6660-9375]{Takao Nakagawa}
    \affiliation{Institute of Space and Astronautical Science, Japan Aerospace Exploration Agency, 3-1-1 Yoshinodai, Chuo-ku, Sagamihara, Kanagawa 252-5210, Japan}
\author[0000-0001-8511-2981]{Norio Narita}
    \affiliation{Komaba Institute for Science, The University of Tokyo, 3-8-1 Komaba, Meguro, Tokyo 153-8902, Japan}
    \affiliation{Astrobiology Center, 2-21-1 Osawa, Mitaka, Tokyo 181-8588, Japan}
    \affiliation{Instituto de Astrof\'{i}sica de Canarias, 38205 La Laguna, Tenerife, Spain}
\author[0000-0001-9326-8134]{Jun Nishikawa}
    \affiliation{National Astronomical Observatory of Japan, 2-21-1 Osawa, Mitaka, Tokyo 181-8588, Japan}
    \affiliation{Astrobiology Center, 2-21-1 Osawa, Mitaka, Tokyo 181-8588, Japan}
    \affiliation{Department of Astronomical Science, The Graduate University for Advanced Studies, SOKENDAI, 2-21-1Osawa, Mitaka, Tokyo 181-8588, Japan}
\author[0000-0002-8300-7990]{Masahiro Ogihara}
    \affiliation{Tsung-Dao Lee Institute, Shanghai Jiao Tong University, 520 Shengrong Road, Shanghai 201210, China}
\author[0000-0002-5051-6027]{Masashi Omiya}
    \affiliation{Astrobiology Center, 2-21-1 Osawa, Mitaka, Tokyo 181-8588, Japan}
    \affiliation{National Astronomical Observatory of Japan, 2-21-1 Osawa, Mitaka, Tokyo 181-8588, Japan}
\author{Takuma Serizawa}
    \affiliation{Institute of Engineering, Tokyo University of Agriculture and Technology, 2-24-16 Nakacho, Koganei, Tokyo 184-8588, Japan}
    \affiliation{National Astronomical Observatory of Japan, 2-21-1 Osawa, Mitaka, Tokyo 181-8588, Japan}
\author{Akitoshi Ueda}
    \affiliation{Astrobiology Center, 2-21-1 Osawa, Mitaka, Tokyo 181-8588, Japan}
    \affiliation{National Astronomical Observatory of Japan, 2-21-1 Osawa, Mitaka, Tokyo 181-8588, Japan}
    \affiliation{Department of Astronomical Science, The Graduate University for Advanced Studies, SOKENDAI, 2-21-1Osawa, Mitaka, Tokyo 181-8588, Japan}
\author{S\'{e}bastien Vievard}
    \affiliation{Subaru Telescope, 650 N. Aohoku Place, Hilo, HI 96720, USA}
\author[0000-0002-4361-8885]{Ji Wang}
    \affiliation{Department of Astronomy, The Ohio State University, 100 W 18th Avenue, Columbus, OH 43210, USA}



\begin{abstract}
The Subaru telescope is currently performing a strategic program (SSP) using the high-precision near-infrared (NIR) spectrometer IRD to search for exoplanets around nearby mid/late-M~dwarfs via radial velocity (RV) monitoring. As part of the observing strategy for the exoplanet survey, signatures of massive companions such as RV trends are used to reduce the priority of those stars. However, this RV information remains useful for studying the stellar multiplicity of nearby M~dwarfs. To search for companions around such ``deprioritized" M~dwarfs, we observed 14 IRD-SSP targets using Keck/NIRC2 observations with pyramid wavefront sensing at NIR wavelengths, leading to high sensitivity to substellar-mass companions within a few arcseconds. We detected two new companions (LSPM~J1002+1459~B and LSPM~J2204+1505~B) and two new candidates that are likely companions (LSPM~J0825+6902~B and LSPM~J1645+0444~B) as well as one known companion. Including two known companions resolved by the IRD fiber injection module camera, we detected seven (four new) companions at projected separations between $\sim2-20$~au in total. A comparison of the colors with the spectral library suggests that LSPM~J2204+1505~B and LSPM~J0825+6902~B are located at the boundary between late-M and early-L spectral types. Our deep high-contrast imaging for targets where no bright companions were resolved did not reveal any additional companion candidates. The NIRC2 detection limits could constrain potential substellar-mass companions ($\sim10-75\ M_{\rm Jup}$) at 10~au or further. The failure with Keck/NIRC2 around the IRD-SSP stars having significant RV trends makes these objects promising targets for further RV monitoring or deeper imaging with JWST to search for smaller-mass companions below the NIRC2 detection limits.

\end{abstract}


\section{Introduction} \label{sec: Introduction}

Stellar multiplicity plays a significant role in the architecture and evolution of stellar and planetary systems. A large fraction of stellar systems are binaries or higher-order multiples \citep[e.g.][]{turner2008,raghavan2010,janson2012}, with orbital semi-major axes spanning from a few stellar radii to thousands of au \citep[e.g.][]{duchene2013}. 
The dynamical influence of stellar companions \citep[e.g.][]{kozai1962,lidov1962,holman1999} can affect the frequency \citep[e.g.][]{ziegler2018,asensio2018} and orbital properties \citep[e.g.][]{winn2009,fontanive2019} of planets in binary systems. From an observational viewpoint, unresolved binarity can lead to systematic errors or uncertainties in the measured properties of stars and planets \citep[e.g.][]{daemgen2009,kopytova2016}. Conversely, a well-characterized stellar binary provides useful astrophysical information that would otherwise be difficult to attain, such as the dynamical masses of stars \citep[e.g.][]{montet2015,calissendorff2022}. Simultaneous knowledge of both the brightness and the dynamical mass of stars is especially important for constraining isochronal and evolutionary models \citep[e.g.][]{baraffe1998,feiden2016}, which remain uncertain particularly for active M-type stars \citep[e.g.][]{pecaut2016,janson2018,asensio2019}.  

For all of the above reasons, it is important to detect and characterize stellar binaries, especially around nearby stars where multiple different techniques (such as radial velocity, astrometry, and imaging) can be applied to gain deeper insights into the system properties. The \textit{Gaia} \citep{Gaia_2016_main} mission, monitoring $\sim 10^9$ stars across the full sky and acquiring dynamical information for all of them, will be highly useful in this regard. However, \textit{Gaia} has limitations for the characterization of close-in and intermediate-orbit binary pairs. 
The \textit{Gaia} telescope generally cannot spatially distinguish binaries with projected separations much smaller than $\sim$1$^{\prime \prime}$ \citep{brandeker2019}. While it can still pick up the astrometric motion of the photocenter of unresolved binaries in many cases, the amplitude of this motion depends both on the mass ratio and the brightness ratio of the stellar pair, thus creating a degeneracy that \textit{Gaia} by itself cannot resolve. 
Thus, high-resolution imaging with adaptive optics (AO) techniques is crucial for acquiring spatially resolved properties of close and intermediate-orbit binaries, and the resulting scientific potential is particularly large for late (M-type) stars. 

The Subaru strategic program employing the high-precision near-infrared (NIR) spectrometer \citep[R$\sim$70,000;][]{Tamura2012,Kotani2018} is performing a large campaign of RV measurements targeting $>120$ nearby ($\lesssim25$~pc) M~dwarfs less massive than $\sim0.3\ M_\odot$ since 2019 (hereafter IRD-SSP; PI: Bun'ei Sato). It has already yielded a new super-Earth around Ross~508 \citep{Harakawa2022}, as well as other candidates.
The program primarily aims to search for exoplanets, particularly terrestrial planets in the habitable zone, and in the cases where we find signatures of massive companions (see Section \ref{sec: Data} for details) that would preclude us from searching for such planets, we stop the monitoring of those targets in the IRD-SSP campaign. 

However, such RV trends are high-quality indications of outer stellar (or sub-stellar) companions, which as we have outlined above are important to identify and characterize with spatially resolved imaging. 
We therefore used Keck/NIRC2 to examine the origin of the observed RV trends with AO imaging.
Adaptive optics using optical wavefront sensing has poor performance on faint, red targets such as M~dwarfs and its performance is insufficient to detect substellar-mass companions located within a few arcseconds of the target star. 
Here we utilize the $H$-band pyramid wavefront sensing at Keck/NIRC2 \citep{Bond2020} for more effective wavefront sensing and higher AO performance than optical wavefront sensing for the IRD-SSP targets so that we can search these RV-trend systems for companions as small as Jovian-mass planets.
While the high-contrast imaging surveys adopting wavefront sensing in optical wavelengths have been carried out so far \citep[e.g.,][]{Bowler_2015_PALMS}, this survey is the first high angular resolution/contrast explorations for companions around cool M~dwarfs with the NIR wavefront sensing technique.
The application of NIR wavefront sensing to the IRD-SSP targets that show RV/astrometric anomalies provides the pilot study to efficiently conduct companion census for nearby cool M~dwarfs. 

\section{Data} \label{sec: Data}

\subsection{Subaru/IRD-SSP Data} \label{sec: Subaru/IRD}
We obtained one-dimensional spectra from the Subaru/IRD raw data using standard data reduction techniques for echelle spectrographs  \citep[e.g.,][]{Kuzuhara_2018_IRD} and applied the pipeline of \citet{Hirano_2020_IRDpipeline} to the extracted spectra for the RV calculations. 
Note that this pipeline produces relative RVs and the absolute RVs are calculated in Section \ref{sec: Target Selection}.
The RV measurements from the IRD-SSP are sensitive to companions with masses comparable to terrestrial planets through to low-mass M dwarfs. 
The IRD RV measurements sometimes indicate long-term linear trends that should be attributed to wide-separation companions as well as curvatures induced by companions moderately separated from their hosts. 
IRD-SSP focuses on the planet survey around single mid-to-late M dwarfs. Stars with signatures of massive companions are deprioritized in the main survey as mentioned in Section \ref{sec: Introduction}.

For IRD observations, the light from a target star is injected into a fiber for the IRD spectrograph by guiding the star with a CCD camera at the fiber injection module (FIM). 
The FIM camera has a sensitivity around $zY$-band (covering wavelengths from 0.83~\micron\ to 1.05~\micron) and provides AO-corrected imaging for the target at each telescope pointing. We investigated the FIM images during the SSP campaign to search for any bright companions around the IRD-SSP targets. However, the FIM data generally have short exposures and the AO performance is limited due to the faintness of the targets for the AO188's optical wavefront sensor, so the FIM data are not sensitive enough to detect substellar-mass companion. 
The AO-corrected images of the FIM camera before injecting lights from targets into a fiber of IRD are available to estimate the centroid of central stars.  
Dark images were subtracted from the raw CCD images to remove background signals, while flat-fielding corrections were not applied to the images because artificial anomaly patterns appear after the corrections. 
We corrected the optical distortion of the FIM camera using the distortion map modeled with the images of the M5 globular cluster (Kuzuhara et al. in prep).
After the distortion corrections, the FIM images have a plate scale of 67.000 $\pm$ 0.023 mas~pixel$^{-1}$.
Because the FIM images in the IRD-SSP were obtained with the pupil-tracking mode, we needed to calculate their parallactic angles to align the y-axis of images to the north direction. 
The calculations were based on the parallaxes, proper motions, right ascension, and declination from {\it Gaia} DR3 for all the targets except those which {\it Gaia} did not detect. 
The FIM camera resolved a new companion and some companions that had been previously resolved \citep{Janson2014,Cortes2017,RoboAO2020} during the IRD-SSP campaign.
The astrometry measurements of the companions revealed in the IRD-FIM images are presented in Section \ref{sec: Subaru/IRD-FIM}.

\subsection{Target Selection for Keck/NIRC2} \label{sec: Target Selection}

Table \ref{tab: target list} presents the target list of the Keck/NIRC2 observations and their properties.
The target selection is based on RV trends appearing in the IRD observations.  
The RV trends were computed by simply fitting linear equations to the RV measurements of our targets. 
We then used {\tt curve\_fit} in {\tt scipy.optimize} with the option of absolute\_sigma = True, which is equivalent to controlling reduced chi squares in the fitting to unity.
We did not include the targets where known companions were resolved by the IRD-FIM camera except for LSPM~J2151+1336, which has a companion at $\sim$0\farcs7 \citep[][]{Cortes2017, RoboAO2020}. As the companion was clearly resolved in the IRD-FIM data (see Section \ref{sec: Subaru/IRD-FIM}) we did not monitor RV but included this system to the NIRC2 target list to check if there were other closer companions that might be contributing to the astrometric perturbation suggested by the RUWE ``goodness-of-fit" parameter included in the {\it Gaia} DR3 catalog \citep{Gaia-EDR3}.
The NIRC2 follow-up observations thus prioritized those with the RV trends and unresolved by the FIM camera. 
Note that part of the unresolved targets with small RV trends have not fully been dropped from the IRD-SSP target list.
We also note that this paper only presents the RV trends indicated by the IRD-SSP campaign. RV measurements from each exposure will be presented in forthcoming papers with detailed characterizations.

We also took into account the {\it Gaia} RUWE parameter \citep{Gaia-EDR3}.
Even in cases where we did not find significant RV trends, we include some targets that have larger RUWE values than 1.4, which means the {\it Gaia} astrometry series cannot be fitted with a single star, to offset the degeneracy of orbital inclination implicit in RV measurements. 
We also searched for wider stellar companions with a 2-arcmin radius search in {\it Gaia} DR3 around each target and did not find any objects that share the parallax and proper motion with our targets. 

We note that we do not identify the age of each system as determining the age of field M~dwarfs is systematically difficult. Instead we assume an age range of 1-10~Gyr in the discussions. 
To justify our assumption, we analyzed the kinematic motions using Banyan $\Sigma$ \citep{Gagne2018} to confirm that our targets do not belong to any young associations or moving groups. We included {\it Gaia} astrometry (RA, Dec, proper motion, and parallax), and median absolute RV (see Table \ref{tab: target list}) from our observations. 
The absolute RVs were measured using a relatively telluric-free wavelength region of 1030--1330 nm.
We took the cross-correlation between the IRD spectra and synthetic spectra based on the VALD line list \citep{1999A&AS..138..119K, 2015PhyS...90e4005R} and the MARCS model atmosphere \citep{2008A&A...486..951G}, and made the barycentric correction.
The errors correspond to the nominal standard errors of the absolute RV values derived from the individual frames.
As mentioned above, we did not measure the RV of LSPM~J2151+1336, and instead we used a literature RV value \citep[$-26.220\pm0.007$~km\ s$^{-1}$;][]{Fouque2018}.
For LSPM~J1534+1800, which {\it Gaia} did not detect, we referred to the UCAC4 catalog \citep{Zacharias2013} for the coordinates and proper motion and \cite{Dittmann_2014_distance} for the parallax.
Eventually the analysis output showed all of our targets are likely field stars ($>5\sigma$).

\startlongtable
\begin{deluxetable*}{ccccccccc}
\centerwidetable
\tablecaption{Properties of the Keck/NIRC2 Targets}
\label{tab: target list}
\tablehead{\colhead{Target} & \colhead{SpType$^a$} & \colhead{Temperature$^a$} & \colhead{Mass$^b$} & \colhead{RV Trend$^c$ (S/N)} & \colhead{Absolute RV$^c$} &  \colhead{RUWE$^d$} & \colhead{Distance$^d$} & \colhead{Known companion$^e$} \\
\colhead{} & \colhead{} & \colhead{K} & \colhead{$M_\odot$} & \colhead{m\ s$^{-1}$ yr $^{-1}$ ($\sigma$)} & \colhead{km~s$^{-1}$} & \colhead{} & \colhead{pc}  & \colhead{}
}

\startdata
 LSPM J0008+4918 & \dots & 2878$^\dagger$ & 0.11 & $-$287 (3.0) & $-2.33\pm0.66$ & 1.1 & 14.76 $\pm$0 .04 & \\ 
 LSPM J0044+0907 & M4 $\pm$ 0.2 & 3160 & 0.36 & $-$12987 (13) & $-8.5\pm2.7$ & 1.6 & 24.99 $\pm$ 0.05 & \\ 
 LSPM J0336+1350 & M4 $\pm$ 0.2 & 3221 & 0.21 & 152 (1.3) & $-3.50\pm0.24$ & 1.9 & 28.84 $\pm$ 0.03 & \\ 
 LSPM J0825+6902 & M6 $\pm$ 0.5 & 2898 & 0.14 & $-$295 (49) & $-5.48\pm0.25$ & 4.6 & 12.29 $\pm$ 0.06 & \\ 
 LSPM J0859+7257 & \dots & 3150$^\dagger$ & 0.17 & $-6.9$ (2.2) & $-0.77\pm0.01$ & 1.5 & 14.394 $\pm$ 0.005 &  \\ 
 LSPM J1002+1459 & M4.5 $\pm$ 0.4 & 3174 & 0.22 & 68 (2.2) & $1.77\pm0.33$ & 4.4 & 17.85 $\pm$ 0.04 & IRD-FIM (candidate)  \\ 
 LSPM J1534+1800 & M5 $\pm$ 0.3 & 3024 & \dots & $-$427 (6.6) & $-36.00\pm0.07$ & 2.6 & 17.09 $\pm$ 0.02 &  \\ 
 LSPM J1645+0444 & M5 $\pm$ 0.2 & 2960 & 0.13 & $-$54 (2.8) & $39.78\pm0.03$ & \dots & 15.7 $\pm$ 1.1 & \\ 
 LSPM J1717+1140 & M5 $\pm$ 0.3 & 2994 & 0.15 & $-$288 (11) & $-60.73\pm0.36$ & 5.7 & 12.39 $\pm$ 0.02 & \\ 
 LSPM J1922+0702 & \dots & 3460$^\dagger$ & 0.22 & $-$0.21 (0.04) & $24.26\pm0.27$ & 1.6 & 10.617 $\pm$ 0.003 & \\ 
 LSPM J2043+0445 & M4.5 $\pm$ 0.6 & 2979 & 0.16 & 194 (9.0) & $-48.25\pm0.23$ & 10.2 & 15.05 $\pm$ 0.04 & \\ 
 LSPM J2151+1336 & M4.5 $\pm$ 0.3 & 3141 & \dots & \dots & $-26.220\pm0.007$  & 2.6 & 18.02 $\pm$ 0.02 & Y, IRD-FIM \\ 
 LSPM J2204+1505 & M5 $\pm$ 0.4 & 2947 & 0.15 & 4864 (29) & $-25.0\pm1.2$ & 25.9 & 23.11 $\pm$ 0.38 & \\ 
 LSPM J2338+3909 & M4 $\pm$ 0.5 & 3303 & 0.27 & $-$8.2 (3.5) & $-0.60\pm0.05$ & 1.4 & 21.13 $\pm$ 0.01 & 
\enddata
\tablecomments{
$^a$ The spectral type and temperature parameters are basically referred to \cite{Koizumi2021} ($T_{\rm SED}$ for temperatures) and the TESS input catalog \citep[with a $\dagger$ symbol, TIC;][]{TIC2019}. 
$^b$ The mass parameters are referred to the TIC. 
$^c$ The RV trends and absolute RVs are estimated from our IRD observations. Note that we refer to \cite{Fouque2018} for the absolute RV of LSPM~J2151+1336 because we did not have IRD-RV measurements.
$^d$ We used {\it Gaia} EDR3 \citep{Gaia-EDR3} to check RUWE and distance except for LSPM~J1645+0444 that has not been recorded in the {\it Gaia} data releases. For the distance of this target we referred to \cite{Dittmann2014}. 
$^e$ `Y' indicates previous studies reported the companion and `IRD-FIM' indicates the IRD-FIM camera resolved the companion. See Sections \ref{sec: Target Selection} and \ref{sec: Subaru/IRD-FIM} for details. We do not include the newly detected companions by the NIRC2 observations in this column.
 }
\end{deluxetable*}

\subsection{Keck/NIRC2 Data} \label{sec: Keck/NIRC2 Data}

We performed Keck/NIRC2 adaptive optics imaging  using pyramid $H$-band wavefront sensing \citep{Bond2020} for precise wavefront correction on our red, mid/late M-type targets \citep[see also the comparison of high-contrast imaging performance in][]{Uyama2020}. 
The observations (PI: Charles Beichman) were allocated on 2022 February 13 (DIMM seeing: $>1\farcs2$), August 08 (seeing: $\sim1^{\prime\prime}$), and November 08 UT (seeing: 0\farcs5-0\farcs6) and the observing log is summarized in Table \ref{tab: obs log}. Note that many of the data sets on 2022 February 13 and August 08 were not taken under good seeing conditions. There were many poor-AO image frames (up to $\sim50\%$) in each obtained data set; we measured amplitude and full width at half maximum (FWHM) of the PSF and consistently removed such poor frames, such that the total integration time in the observing log represents only good AO data sets.

As an initial step, we used the broad-band $K_{\rm s}$/$K^\prime$ filters or the narrow-band Br$\gamma$ filter to search for bright companions. On 2022 November 08, the AO performance was better than the other dates and the core of the point spread function (PSF) could easily be saturated with the broad-band filters. Then we instead used the Br$\gamma$ filter.
For the targets where we did not find any bright companion candidates by eye during the observations, we conducted deep $L^\prime$ observations to search for faint and cold companions utilizing angular differential imaging \citep[ADI;][]{Marois2006}. When we tried using the coronagraphic mask, speckles outside the coronagraph were not seen, making it difficult to align the PSF with the center of the coronagraph. Therefore, we did not use the coronagraph in the bulk of our observations.
Note that we did not conduct deep $L^\prime$-band observations for LSPM~J1534+1800 and LSPM~J0336+1350, despite the null detection with the $K^\prime$ and Br$\gamma$ filters respectively (see Section \ref{sec: Results}).
This is because we did not have sufficient time to conduct deep imaging of that target on the allocated nights. 
We tested the $L^\prime$-band exposures when observing LSPM~J0859+7257 (on 2022 February 13), LSPM~J1645+0444 (on August 08), and LSPM~J2204+1505 (on November 08) that likely harbor companions and used these imaging data sets for checking the color of these companions resolved in this survey (see Section \ref{sec: Results}).

\begin{deluxetable*}{ccccccccc}
\tablecaption{Keck/NIRC2 Observing Log}
\label{tab: obs log}
\tablehead{ \colhead{Target} & \colhead{Filter} & \colhead{Magnitude$^a$} & \colhead{$t_{\rm int}$$^b$ [sec]} & \colhead{N$_{\rm exp}$$^c$} & \colhead{FWHM [pix]} & \colhead{Reduction} & \colhead{Field Rotation$^d$} & \colhead{Detection$^e$}}
\startdata
\multicolumn{9}{c}{2022 February 13} \\
 LSPM J0825+6902 & $K^\prime$ & $9.24\pm0.01$ & 5 & 6 & 8.3 & standard & \dots & Y \\ 
  & $L^\prime$ & 8.67 $\pm$ 0.03 & 3 & 1 & 9.9 & standard & \dots & Y \\ 
 LSPM J0859+7257& $L^\prime$ & 8.52 $\pm$ 0.03 & 30 & 81 & 10.3 & ADI & 32\fdg7 & N \\ 
 LSPM J1002+1459 & $K_{\rm s}$ & 8.83 $\pm$ 0.02 & 7.5 & 5 & 7.8 & standard & \dots & Y \\  
\hline \multicolumn{9}{c}{2022 August 08} \\
 LSPM J1534+1800 & $K^\prime$ & 9.79 $\pm$ 0.02 & 3 & 5 & 5.4 & standard & \dots & N  \\ 
 LSPM J1645+0444 & $K^\prime$ & 9.87 $\pm$ 0.02 & 3 & 8 & 5.9 & standard & \dots & Y \\ 
  & $L^\prime$ & 9.36 $\pm$ 0.03 & 15 & 5 & 10.3 & standard & \dots & Y \\ 
 LSPM J1717+1140 & $L^\prime$ & 8.51 $\pm$ 0.03 & 10.5 & 70 & 9.1 & ADI & 16\fdg2 & N \\ 
 LSPM J1922+0702& $L^\prime$ & 7.29 $\pm$ 0.04 & 12 & 35 & 7.4 & ADI & 19\fdg4 & Y \\ 
 LSPM J2043+0445& $L^\prime$ & 8.70 $\pm$ 0.03 & 24 & 65 & 9.4 & ADI & 55\fdg7 & N \\ 
 LSPM J2151+1336& $K^\prime$ & 8.50 $\pm$ 0.02 & 4 & 2 & 7.7 & standard & \dots & Y \\ 
 LSPM J2204+1505& $K^\prime$ & 10.06 $\pm$ 0.02 & 4 & 3 & 5.9 & standard & \dots & Y \\ 
\hline \multicolumn{9}{c}{2022 November 08} \\
 LSPM J0008+4918 & $L^\prime$ & 9.54 $\pm$ 0.03 & 30 & 26 & 8.2 & ADI & 16\fdg8 & N \\ 
 LSPM J0044+0907 & $L^\prime$ & 8.23 $\pm$ 0.03 & 28 & 25 & 8.1 & RDI$^f$ & 6\fdg0 & N \\ 
 LSPM J0336+1350 & Br$\gamma$ & 10.08 $\pm$ 0.02 & 5 & 22 & 5.1 & standard & \dots & N \\ 
 LSPM J1922+0702& Br$\gamma$ & 7.77 $\pm$ 0.02 & 50 & 50 & 5.0 & standard & \dots & Y \\ 
 LSPM J2204+1505& $L^\prime$ & 9.52 $\pm$ 0.03 & 27 & 15 & 7.3 & standard & \dots & Y \\ 
 LSPM J2338+3909& $L^\prime$ & 8.35 $\pm$ 0.03 & 28 & 30 & 8.0  & ADI & 14\fdg8 & N \\ 
\enddata
\tablecomments{$^a$ The apparent magnitude is converted from 2MASS $K_{\rm s}$ photometry (for the NIRC2 $K_{\rm s}$, $K^\prime$, and Br$\gamma$ filters) or interpolation of WISE W1 and W2 photometry (for the $L^\prime$ filter) using {\tt species} \citep{species}. For the targets with bright companions, both primary and secondary are integrated into the magnitude parameter. $^b$ The product of the single exposure time per coadd and the number of coadds per exposure. $^c$ The poor-AO data frames were not counted in the number of exposures. $^d$ If ADI was applied. $^e$ Detection of any companion candidates including the known companions. $^f$ The field rotation was insufficient to conduct ADI reduction.}
\end{deluxetable*}

For all the data sets, we conducted standard calibration; dark ($K_{\rm s}$ and $K^\prime$ data taken on 2022 February 13) or sky subtraction (other data), flat fielding, and bad-pixel masking, and then we corrected for distortion \citep{Yelda2010, Service2016} and aligned the PSF of the central star. 
For the shallow-imaging data sets we derotated the images to align North upward and combined them, and checked the brightness  and location of any companion candidates.

For the deep-imaging $L^\prime$ data sets we then proceeded with an ADI reduction \citep{Marois2006} where a reference PSF that incorporates much of the quasi-static speckle noise can be produced by taking into account the fact that the speckles stay fixed to the telescope pupil as the FoV rotates. 
We used {\tt pyklip} \citep{pyklip} for post-processing, which produces the most likely reference PSFs based on the Karhunen-Lo\`eve Image Processing algorithm  \citep[KLIP;][]{Soummer2012}. To explore faint companions around the targets without bright companion candidates, we divided the FoV into 5 annular areas (annuli: 5 in the {\tt pyklip} settings) and adopted Karhunen-Lo\`eve (KL) mode=20 and movement=1, which determines the aggressiveness of the PSF subtraction, for the final output of the ADI data sets. 
The LSPM~J2338+3909 and LSPM~J0044+0907 data sets obtained smaller field rotations than other deep imaging data sets (14\fdg8 and 6\fdg0, respectively). For the LSPM~2338+3909 data set we adopted a moderate KL mode of 10 and used the whole FoV (annuli: 1) to make the reference PSF for ADI. 
For the LSPM~J2338+3909 data set the field rotation was insufficient to conduct ADI. We instead utilized other deep $L^\prime$-band data on the same date as a PSF reference library and conducted reference-star differential imaging \citep[RDI;][]{Ruane2019}. We included the science (LSPM~J0044+0907) and the reference (LSPM~J0008+4918 and LSPM~J2338+3909) frames in the PSF library. We then calculated the mean square errors \citep{Wang2004} of the reference frames at separations $\lesssim 5\lambda/D$ and removed the worst 20\% frames from the PSF library. Finally we conducted RDI-based PSF subtraction using the {\tt pyklip} algorithms, where we adopted the KL mode of 20 and annuli of 1.

On 2022 November 08 we conducted deep Br$\gamma$ imaging with LSPM~J1922+0702 to investigate the color of the companion candidates detected on 2022 August 08. We did not conduct ADI for this follow-up observation because these candidates are located well outside the stellar halo (see Section \ref{sec: Results}).
The data reduction follows those on the shallow imaging data sets. After combining the whole set of exposures there remained hot pixels which were masked out to yield a final image and to calculate properly the signal-to-noise ratios (SNRs) of the companion candidates.

\section{Results} \label{sec: Results}

\subsection{Subaru/IRD-FIM} \label{sec: Subaru/IRD-FIM}

\begin{figure*}
    \centering
    \includegraphics[width=0.9\textwidth]{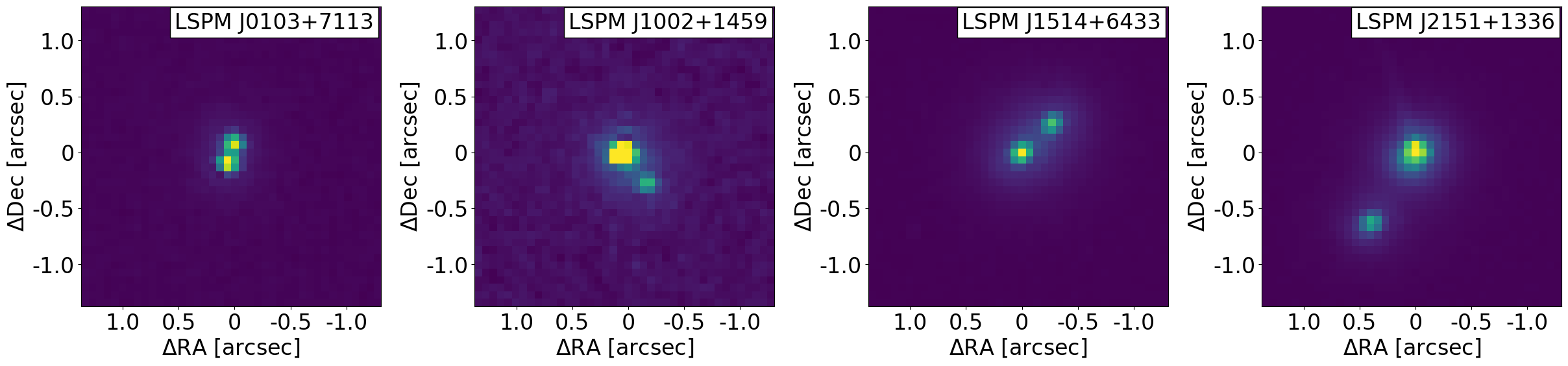}
    \caption{Summary of the Subaru/IRD-FIM images that resolved companions. The target name is labeled at the upper right of each panel. All the images are rotated such that north up and the primary star is located at the center. The color scale is arbitrarily set to clearly show our detections.}
    \label{fig: summary IRD-FIM}
\end{figure*}

We present our IRD-FIM results that resolved the companions in Figure \ref{fig: summary IRD-FIM} and Table \ref{tab: target list - FIM}.
We detected a new companion candidate around LSPM~J1002+1459 and the previously resolved companion candidate around LSPM~J1514+6433 \citep{RoboAO2020} as well as the confirmed known stellar companions - LSPM~J0103+7113~B \citep{Janson2014} and LSPM~J2151+1336~B \citep[][and see also our NIRC2 result in Section \ref{sec: Results}]{Cortes2017,RoboAO2020}.
We note that the sensitivity dependency on the wavelength of the IRD-FIM CCD has not been well characterized and we thus present only astrometric information in this study.

\begin{figure}
    \centering
    \includegraphics[width=0.45\textwidth]{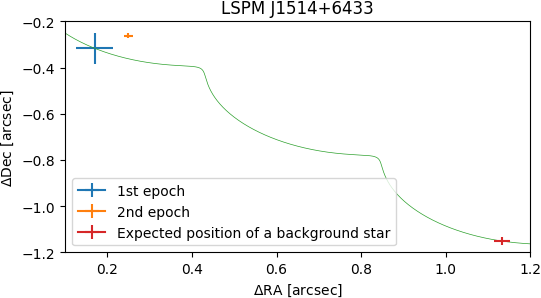}
    \caption{Common proper motion test of the companion candidate around LSPM~J1514+6433. The first and second epochs correspond to the Robo-AO observation \citep{RoboAO2020} and the IRD-FIM observation dates respectively. The expected position corresponds to the simulated position assuming a background star with a zero proper motion on the IRD observation date based on the first position. The green line indicates the trajectory of the simulated motion of the background star. We also assumed a zero proper motion acceleration for the primary star induced by the companion.}
    \label{fig: cpm test J1514}
\end{figure}

For LSPM~J1514+6433 and LSPM~J2151+1336, we fitted an elliptical Gaussian to the PSFs of the central stars and the companion candidates to determine the centers of each PSF. 
LSPM~J0103+7113 and LSPM~J1002+1459 are tight binaries and fitting with elliptical Gaussian functions did not work well, then we applied reference PSFs of single stars observed after/before these targets to the PSF fitting.
The uncertainties of astrometry measurements include the errors in the distortion modeling.
While \cite{RoboAO2020} did not present a proper motion test for the companion candidate around LSPM~J1514+6433, the combination of our FIM measurements with the previous astrometric data confirmed that this candidate is a real companion (see Figure \ref{fig: cpm test J1514}).
We note that simulating the motion of the background star is based on the proper motion of the primary star. Given that our targets show RV trends, they might have proper motion accelerations. This could be a potential uncertainty in simulating the background motion but this effect cannot be simply evaluated, particularly when we do not detect companions. Therefore we assume a zero proper motion acceleration for the primary star.

\startlongtable
\begin{deluxetable*}{ccccc|ccc}
\tablecaption{Properties of the IRD-SSP Targets resolved by the IRD-FIM Data}
\label{tab: target list - FIM}
\tablehead{\multicolumn{5}{c}{Primary$^a$} & \multicolumn{3}{c}{Companion}\\
\colhead{Target} & \colhead{SpType} & \colhead{Temperature [K]} & \colhead{Mass [{$M_\odot$}]} & \colhead{Distance [pc]} & \colhead{MJD} & \colhead{Separation} & \colhead{Position Angle} }
\startdata
 LSPM J0103+7113 & \dots & 3261$^\dagger$ & 0.35 & 27.53 $\pm$ 0.07 & 58679.578 & $0\farcs171\pm0\farcs010$ & $343\fdg5 \pm 3\fdg3$ \\ 
 LSPM J1002+1459 & M4.5 $\pm$ 0.4 & 3174 & 0.22 & 17.85 $\pm$ 0.04 & 58564.472 & $0\farcs327 \pm 0\farcs011$  & $221\fdg7 \pm 2\fdg1$ \\
 LSPM J1514+6433 & M3.5$^b$ & \dots & 0.20  & 17.48 $\pm$ 0.02$^c$  & 58652.418 & $0\farcs362 \pm 0\farcs008$ & $313\fdg6 \pm 1\fdg2$ \\ 
 LSPM J2151+1336 & M4.5 $\pm$ 0.3 & 3141 & \dots & 18.02 $\pm$ 0.02 & 58652.531 & $0\farcs743 \pm 0\farcs016$ & $147\fdg9 \pm 1\fdg3$ 
\enddata
\tablecomments{
$^a$ The formatting for the primary is the same as Table \ref{tab: target list} except for the RV trend and RUWE. $^b$ We referred to \cite{Alonso-Floriano2015}. $^c$ {\it Gaia} does not record the parallax for this system and we instead adopted the distance from \cite{Dittmann_2014_distance}.
 }
\end{deluxetable*}

\subsection{Keck/NIRC2} \label{sec: Keck/NIRC2 result}

\begin{figure*}
    \centering
    \includegraphics[width=0.73\textwidth]{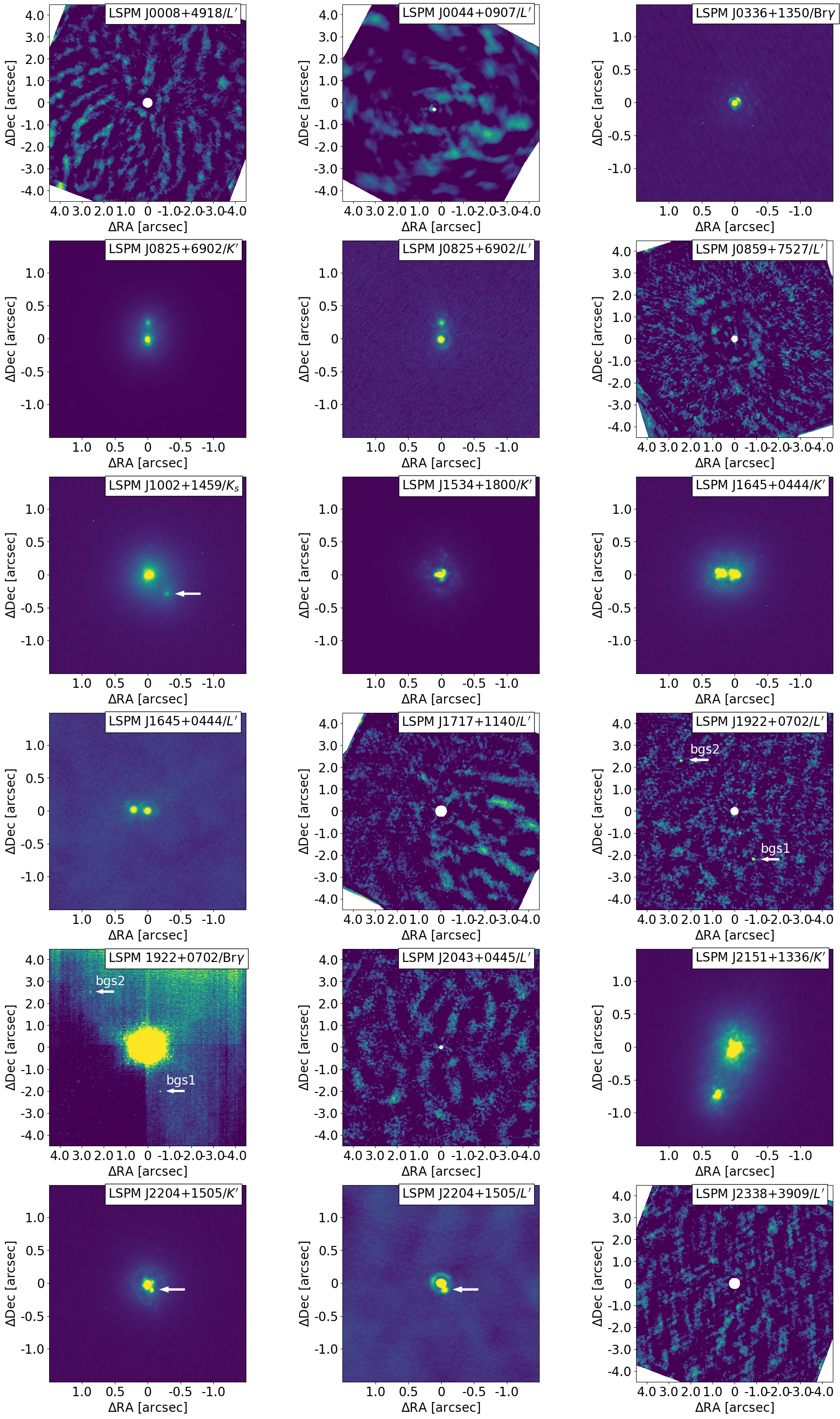}
    \caption{Summary of the Keck/NIRC2 observations. The target name and used filter are labeled at the upper right of each panel. All the images are rotated such that north up and the primary star is located at the center. The color scale is arbitrarily set to clearly show our detections. The ADI/RDI post-processed panels (LSPM~J0008+4918, LSPM~J0044+0907, LSPM~J0859+7257, LSPM~J1717+1140, LSPM~J1922+0702 ($L^\prime$), LSPM~J2043+0445, and LSPM~J2338+3909) correspond to SNR maps after smoothing the output images by a Gaussian ($\sigma$=3~pix). For these data sets the central star is masked by the algorithm. The faint companion candidates are indicated by white arrows and detailed information about the detected companion candidates is summarized in Table \ref{tab: companion candidates}.}
    \label{fig: summary result}
\end{figure*}

We confirmed the known companion around LSPM~J2151+1336, where we did not detect any other close-in companion candidates, and detected six new companion candidates including the new candidate around LSPM~J1002+1449 resolved by the IRD-FIM camera.
Figure \ref{fig: summary result} and Table \ref{tab: companion candidates} summarize our observational results. 
The bright companion candidates detected in the NIRC2 data without ADI reductions are located close to the central star. To take into account the stellar flux, we conducted the aperture photometry (aperture radius is equivalent to half-width at half maximum of the central star's PSF) at the same separation but different position angles ranging from +30 to +330 degrees with a 5-degree step compared to the position angle of the companion candidate. We subtracted the average value of these photometries, as the contamination of the stellar halo, from the photometry at the companion candidates and defined the standard deviation as the noise.
For the faint companion candidates around LSPM~J1922+0702, we used {\tt pyklip} modules to produce the SNR map after smoothing the post-processed image with a Gaussian ($\sigma$=3~pix) and adopted the corresponding SNR in Table \ref{tab: companion candidates}. To calculate the flux attenuation and astrometric errors induced by the self-subtraction due to the ADI post-processing, we injected fake sources and reran the {\tt pyklip} algorithms. 
The astrometric error bars include the PSF fitting errors of both the central star and the companion candidate. 
Those of the companion candidates were determined by the averaged errors of the centroid measurements of fake sources.
Each error was estimated by the square root of the diagonal elements of the covariance matrices produced in the least square fitting. The distortion systematic errors are of order 0.1 pix \citep[pix scale: 9.971 mas/pix;][]{Service2016}, which is negligible compared to the astrometric errors.
The contrast measurements in Table \ref{tab: companion candidates} are also corrected by the attenuation ratios.

\begin{deluxetable*}{cccccccc}
\tablecaption{Detected Companion Candidates with NIRC2}
\label{tab: companion candidates}
\tablehead{\colhead{Target} & \colhead{MJD} & \colhead{Filter} & \colhead{SNR} & \colhead{Contrast} & \colhead{Separation} & \colhead{Position Angle} & \colhead{remarks}}
\startdata
LSPM J0825+6902 & 59623.286 & $K^\prime$ & 29.9 &  0.363 & 0\farcs239 $\pm$ 0\farcs004 & 359\fdg25$\pm$0\fdg82 & new candidate  \\
 & 59623.274 & $L^\prime$ & 28.9 & 0.410 & 0\farcs239 $\pm$ 0\farcs002 & 359\fdg01 $\pm$ 0\fdg53 &  \\
LSPM J1002+1459 & 59623.394 & $K_{\rm s}$ & 20.7 & 0.196 & 0\farcs406 $\pm$ 0\farcs003 & 225\fdg21 $\pm$ 0\fdg64 & new companion \\
LSPM J1645+0444 & 59800.280 & $K^\prime$ & 72.7 & 0.930 & 0\farcs219 $\pm$ 0\farcs002 & 84\fdg53 $\pm$ 0.51 & new candidate \\
 & 59800.284 & $L^\prime$ & 29.6 & 0.865 & 0\farcs216 $\pm$ 0\farcs003 & 84\fdg76 $\pm$ 0\fdg61 &\\
LSPM J1922+0702 & 59800.342 & $L^\prime$ & 8.3 & 5.1$\times10^{-4}$ & 2\farcs340 $\pm$ 0\farcs006 & 201\fdg39 $\pm$ 0\fdg66 & bgs1 \\
 & 59891.210 & Br$\gamma$ & 7.7 & 4.8$\times10^{-4}$ & 2\farcs104 $\pm$ 0\farcs004 & 196\fdg06 $\pm$ 0\fdg.88 & bgs1\\
 & 59800.342 & $L^\prime$ & 6.7 & 3.7$\times10^{-4}$ & 3\farcs358 $\pm$ 0\farcs004 & 46\fdg75 $\pm$ 0\fdg75  & bgs2 \\
 & 59891.210 & Br$\gamma$ & 4.4 & 3.9$\times10^{-4}$ & 3\farcs630 $\pm$ 0\farcs007 & 45\fdg96 $\pm$ 1.32 & bgs2 \\
LSPM J2151+1336 & 59800.367 & $K^\prime$ & 116.7 & 0.406 & 0\farcs755 $\pm$ 0\farcs005 & 160\fdg36 $\pm$ 0\fdg97 & known companion\\
LSPM J2204+1505 & 59800.363 & $K^\prime$ & 10.4 & 0.14$^a$ & 0\farcs114 $\pm$ 0\farcs005 & 208\fdg22 $\pm$ 0\fdg93 & new companion \\
 & 59891.264 & $L^\prime$ & 11.8 & 0.153 & 0\farcs115 $\pm$ 0\farcs004 & 205\fdg73 $\pm$ 0\fdg79 & 
\enddata
\tablecomments{$^a$ The PSF core of the primary star is slightly above the linearity limit of 10k counts/coadd. Therefore the primary star's photometry may be underestimated and thus the contrast may be overestimated.}
\end{deluxetable*}

\begin{figure*}
    \centering
    \includegraphics[width=0.9\textwidth]{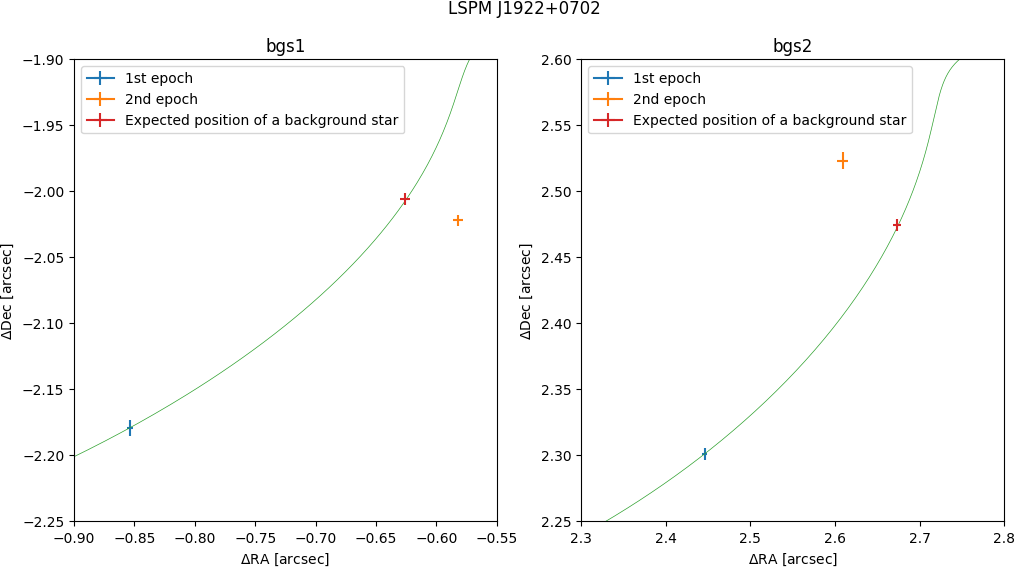}
    \caption{The same proper motion test as Figure \ref{fig: cpm test J1514} for the companion candidates around LSPM~J1922+0702. The 1st and 2nd epochs correspond to the NIRC2 observations on 2022 August 8 and 2022 November 8 respectively. We attribute the difference between the expected position (red cross) and the 2nd epoch (orange cross) to the proper motion of the background stars and potential proper motion acceleration by an unseen companion. }
    \label{fig: cpm test IRD110}
\end{figure*}

\begin{figure*}
\begin{tabular}{cc}
\begin{minipage}{0.5\hsize}
    \centering
    \includegraphics[width=0.9\textwidth]{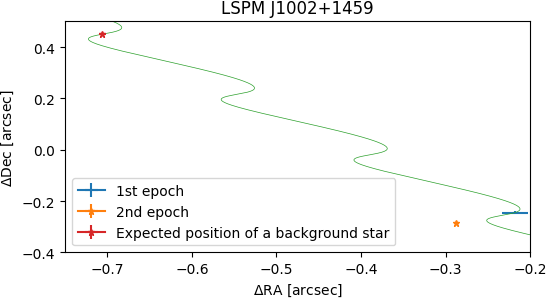}
\end{minipage}
\begin{minipage}{0.5\hsize}
    \centering
    \includegraphics[width=0.9\textwidth]{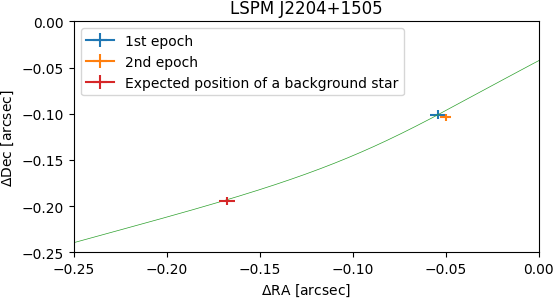}
\end{minipage}
\end{tabular}
\caption{The same proper motion test as Figure \ref{fig: cpm test J1514} for the companion candidates around LSPM~J1002+1459 (left, the 1st and 2nd epochs correspond to the date when the IRD-FIM image was taken and the NIRC2 observations on 2022 February 13 respectively) and LSPM~J2204+1505 (right, the 1st and 2nd epochs correspond to the NIRC2 observations on 2022 August 8 and 2022 November 8 respectively). In the left panel, the astrometric errors of the 2nd epoch and the expected position in case of a background star turn out very tiny in the plot, therefore we instead use star symbols for clarity.}
\label{fig: cpm test IRD56 and 118}
\end{figure*}

Regarding the candidates around LSPM~J1922+0702 and LSPM~J2204+1505, we obtained two epochs of astrometry with NIRC2 and performed common proper motion tests to investigate if they are bound to the host stars. 
We also combined the IRD-FIM result with the NIRC2 result of the companion candidate LSPM~J1002+1459 for the common proper motion test.
Thanks to the high proper motions of the target stars and the precise astrometry by the NIRC2 observations, we concluded that the two companion candidates around LSPM~J1922+0702 are likely background stars with non-zero proper motions (bgs1 and bgs2, see Figure \ref{fig: cpm test IRD110}) and that the companion candidates around LSPM~J1002+1459 and LSPM~J2204+1505 are real companions with a $>45\sigma$ confidence level (see Figure \ref{fig: cpm test IRD56 and 118}). 

As previous observations have not resolved the companions detected by the Keck/NIRC2 observations, the reported photometry should include the companions' fluxes unless the candidates are background objects.
We took into account the contrast ratio between the primary star and the companion, and Table \ref{tab: corrected photometry} presents the corrected photometry of such systems. Note that we used the contrast ratio (and the SNR of the companion candidates) to identify each component, the error bars of the primary stars are dominated by those of the companion candidates.
We then calculated false alarm probabilities (FAPs) of the remaining two companion candidates around LSPM~J0825+6902 and LSPM~J1645+0444 compared with 2MASS catalog \citep{Cutri2003} to compute how likely a background star contaminates the NIRC2 output images. 
We searched for objects located near the target star (radius: $\leq$ 15 arcmin) with an equal or brighter magnitude as the companion candidate, and calculated the number density of such objects. We defined the FAP as the expected number of objects within the separations of the companion candidate from their host stars (see Table \ref{tab: corrected photometry}). 
The FAPs are very small and correspond to $>4\sigma$ confidence level suggesting a real companion. Therefore these companion candidates are statistically likely to be bound to the central stars and we regard these two candidates as companions in this study. The conclusions including the proper motion tests will be presented in future papers with the follow-up observations. As mentioned in Section \ref{sec: Target Selection}, future papers will also present the IRD RV measurements for these companions including orbital fitting results.

The detection frequency from only the NIRC2 observations is 5/14$\sim36\%$, which is a highly biased value from our target selection (see Section \ref{sec: Target Selection}). Thus we do not simply compare it with demographic studies \citep[e.g.][]{Schlecker2022} but we discuss the distributions in terms of mass ratio and separation in Section \ref{sec: Resolved Companions}.
Including the other two companions resolved by the IRD-FIM camera (see Section \ref{sec: Subaru/IRD-FIM}), we detected seven companions in total at projected separations between $\sim2-20$~au.

\begin{deluxetable*}{ccccc}
\tablecaption{Corrected Apparent Magnitude}
\label{tab: corrected photometry}
\tablehead{\colhead{Target} & \colhead{Filter} & \colhead{mag$_{\rm pri}$} & \colhead{mag$_{\rm candidate}$} &\colhead{FAP$^a$}}
\startdata
LSPM J0825+6902 & $K^\prime$ & 9.57 $\pm$ 0.04 & 10.68 $\pm$ 0.04 & 8.5$\times10^{-7}$  \\ 
 & $L^\prime$ & 9.04 $\pm$ 0.05 & 10.01 $\pm$ 0.05 & \dots \\
LSPM J1002+1459 & $K_{\rm s}$ & 9.02 $\pm$ 0.05 & 10.79 $\pm$ 0.05 & N/A \\
LSPM J1645+0444 & $K^\prime$ & 10.58 $\pm$ 0.02 & 10.66 $\pm$ 0.02 & 1.6$\times10^{-6}$ \\ 
 & $L^\prime$ & 10.04 $\pm$ 0.05 & 10.19 $\pm$ 0.05 & \dots \\
LSPM J1922+0702 & Br$\gamma$ & 7.77 $\pm$ 0.02 & 16.07 $\pm$ 0.13 (bgs1), 16.29 $\pm$ 0.22 (bgs2) & N/A \\
 & $L^\prime$ & 7.29 $\pm$ 0.04 & 15.52 $\pm$ 0.13 (bgs1), 15.87 $\pm$ 0.16 (bgs2) & N/A  \\
LSPM J2151+1336 & $K^\prime$ & 8.87 $\pm$ 0.02 & 9.85 $\pm$ 0.02 & N/A \\
LSPM J2204+1505 & $K^\prime$ & 10.20 $\pm$ 0.10 & 12.34 $\pm$ 0.10 & N/A  \\ 
 & $L^\prime$ & 9.67 $\pm$ 0.09 & 11.71 $\pm$ 0.09  & N/A 
\enddata
\tablecomments{$^a$ False alarm probability of a background object using the 2MASS $K_{\rm s}$-band catalog search. We did not calculate the FAPs for the real companions (LSPM J1002+1459, LSPM~J2151+1336, and LSPM~J2204+1505) and the background stars (LSPM~J1922+0702).}
\end{deluxetable*}

\section{Discussion} \label{sec: Discussions}
\subsection{Resolved Companions} \label{sec: Resolved Companions}
\begin{figure*}
\begin{tabular}{cc}
\begin{minipage}{0.5\hsize}
    \centering
    \includegraphics[width=0.9\textwidth]{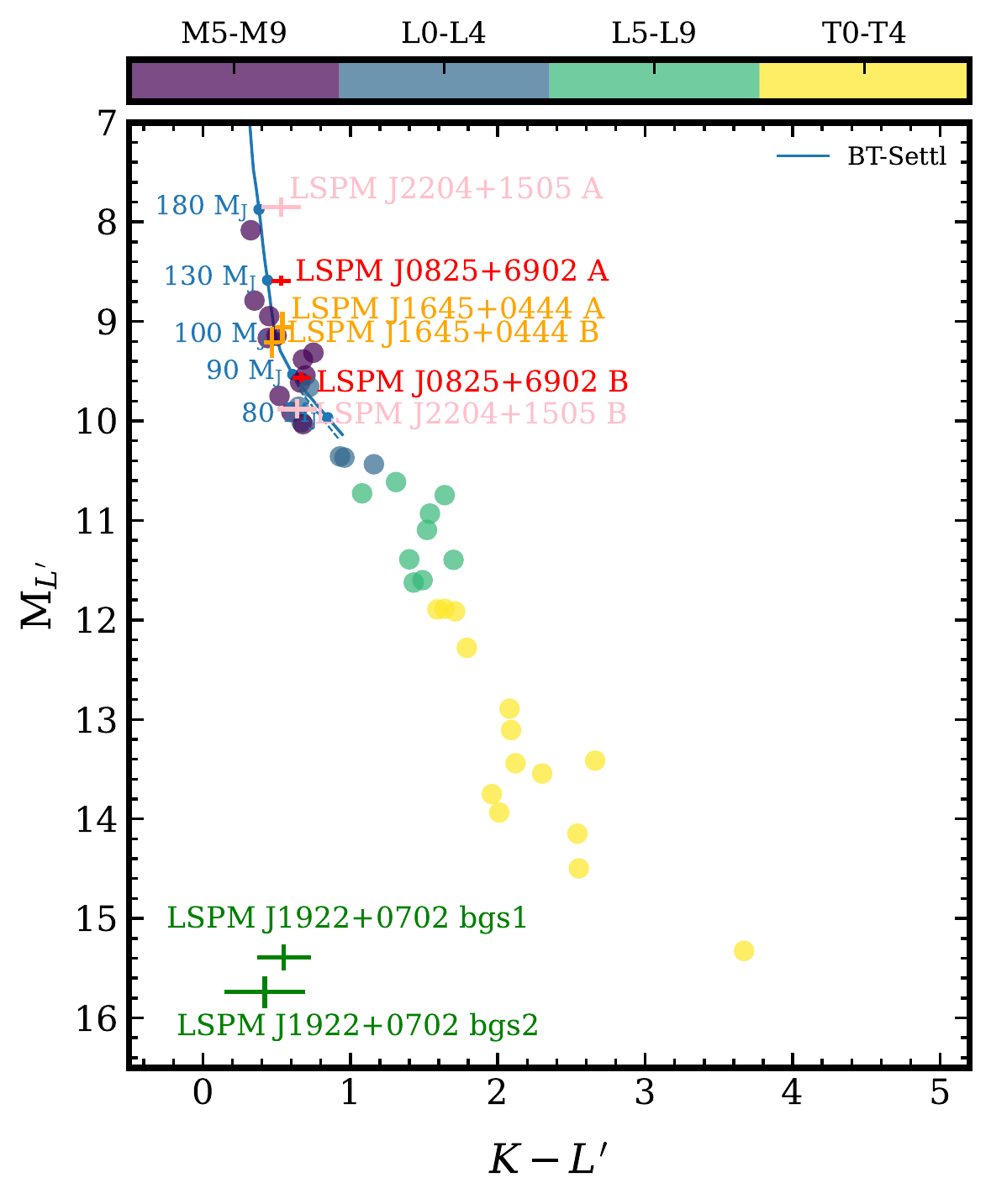}
\end{minipage}
\begin{minipage}{0.5\hsize}
    \centering
    \includegraphics[width=0.9\textwidth]{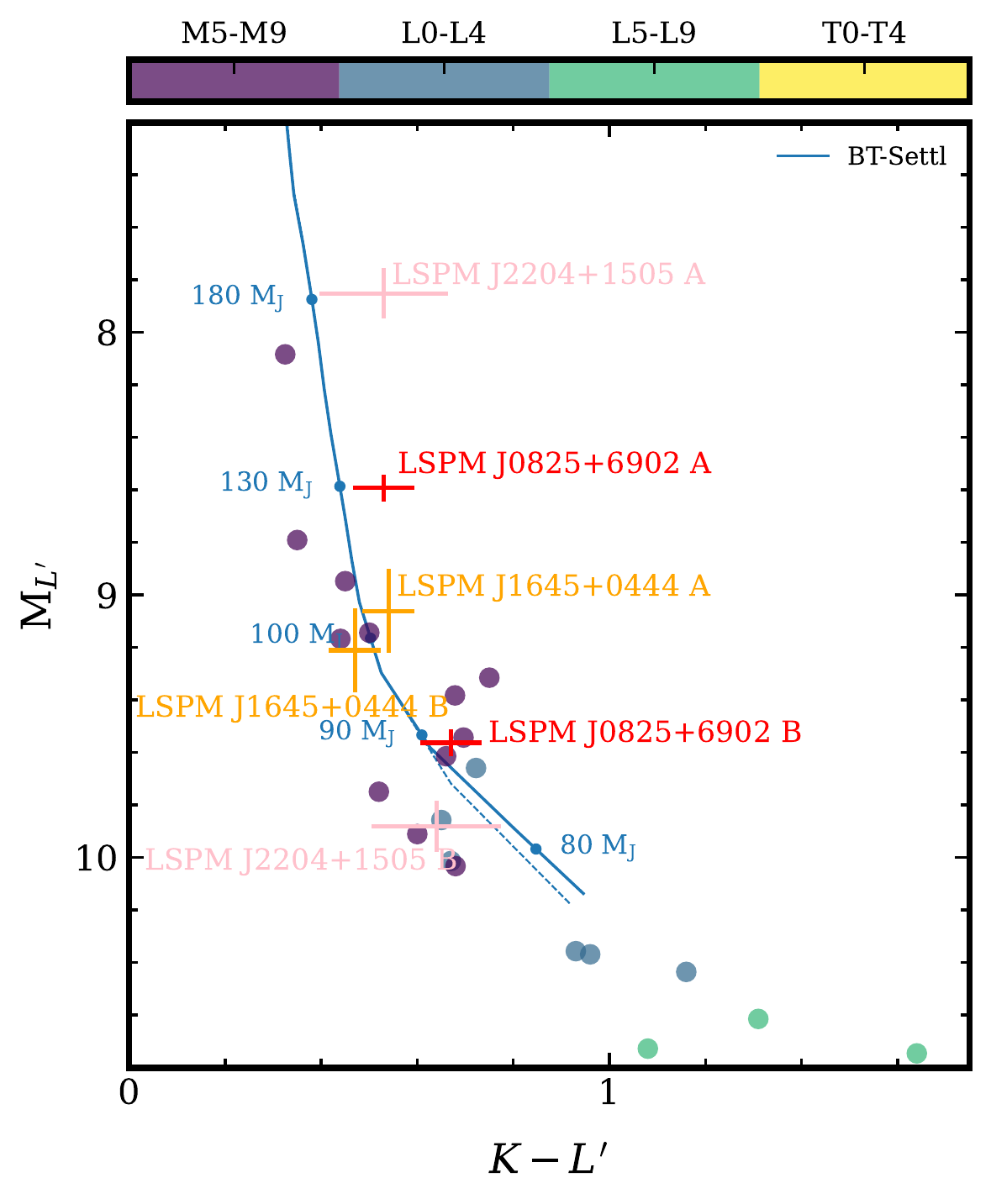}
\end{minipage}
\end{tabular}
\caption{Color-magnitude diagram of the detected companion candidates (we expediently add a suffix of 'B' of the primary star 'A') overlaid with photometric libraries of field low-mass stars/substellar-mass objects \citep{Dupuy2012,Dupuy2013} as well as BT-Settl isochrones \citep[cerulean blue solid: 1~Gyr, dashed: 5~Gyr, and dotted lines: 10~Gyr;][]{Allard2012} and blackbody radiation (gray dashed line). We corrected the Keck/NIRC2 magnitudes into MKO/NSFCam filters to compare with the libraries using {\tt species} \citep{species}. We also added the expected $L^\prime$-band absolute magnitudes of the background stars around LSPM~J1922+0702 (bgs1 and bgs2) assuming the same distance as the target star in the left panel. The right panel shows the zoomed-in diagram without the background stars. }
    \label{fig: color}
\end{figure*}

We resolved the companions around LSPM~J0825+6902, LSPM~J1645+0444, and LSPM~J2204+1505.
Figure \ref{fig: color} shows a color-magnitude diagram of the detected companions that were observed at two filters as well as the primary stars in Table \ref{tab: corrected photometry}. 
The colors of the companions (labeled 'B' in Figure \ref{fig: color}) are consistent with low-mass field dwarfs in terms of spectral libraries \citep{Dupuy2012,Dupuy2013} and an atmospheric model \citep[BT-Settl;][]{Allard2012}, favoring the low-mass companion scenario. In this analysis we set a range of the age between 1~Gyr and 10~Gyr but the BT-Settl isochrones are not sufficiently distinctive to help infer the age of each system from the color. 
LSPM~J0825+6902~B and LSPM~J2204+1505~B are located at the boundary between late-M and early-L spectral types. In particular, LSPM~J2204+1505~B is possibly a massive brown-dwarf companion around a mid-M dwarf. Follow-up spectro-photometric observations and monitoring RV/relative-astrometry will be able to reveal their spectral types and dynamical masses in detail. 
LSPM~J1002+1459 was observed with only the $K_{\rm s}$ filter by Keck/NIRC2 and we have not obtained the $K-L^\prime$ color, but the $K_{\rm s}$-band magnitude is similar to the other companions and consistent with a low-mass object ($\sim100 M_{\rm Jup}$) compared with the atmospheric model. 

We also added the expected $L^\prime$-band absolute magnitudes of the background stars around LSPM~J1922+0702 in Figure \ref{fig: color} assuming the same distance as the target star. These colors are different from very low-mass objects favoring background stars with larger distances than the assumption above.

\begin{figure}
    \centering
    \includegraphics[width=0.47\textwidth]{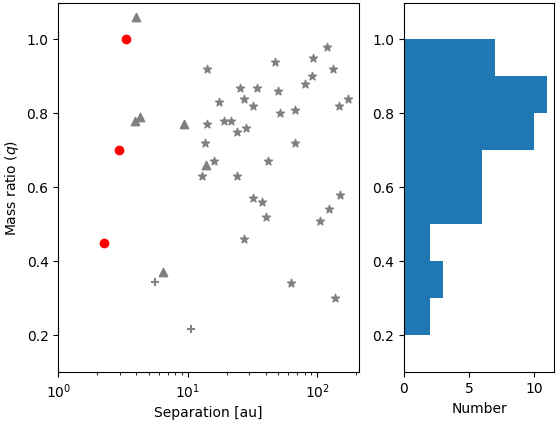}
    \caption{Left: Mass ratio ($q$) vs separation diagram of the resolved companions in this study (red; only those whose mass ratios we estimated) overlaid with other imaging studies (gray). The star, triangle, and plus symbols correspond to \cite{DeFurio2022b,DeFurio2022a}, \cite{calissendorff2022}, and \cite{Biller2022}, respectively. 
    We adopted the semi-major axes derived from orbital fitting for the companions in \cite{calissendorff2022} and \cite{Biller2022} while the projected separations for the resolved companions in this study and \cite{DeFurio2022b, DeFurio2022a}. Right: Histogram of the mass ratios of the companions in the left panel. }
    \label{fig: mass ratio}
\end{figure}

The rough mass ratios, which are estimated from the comparison with the Ames-COND atmospheric model, between the primary and the companion are $q\sim$0.7, 1, and 0.45 for LSPM~J0825+6902 ($M_{\rm pri}\sim130M_{\rm Jup}$, $M_{\rm sec}\sim90M_{\rm Jup}$), LSPM~J1645+0444 ($M_{\rm pri}\sim M_{\rm sec}\sim100M_{\rm Jup}$), and LSPM~J2204+1505 ($M_{\rm pri}\sim180M_{\rm Jup}$, $M_{\rm sec}\sim80M_{\rm Jup}$), respectively. 
Figure \ref{fig: mass ratio} shows a mass-ratio ($q$) vs separation diagram of the resolved companions in our study compared with previous imaging studies, and a histogram of the mass ratio. 
Note that the mass range of the primary star is different in each study; $M_{\rm pri}\lesssim0.3 M_\odot$ in this study, $M_{\rm pri}\lesssim 0.15 M_\odot$ in \cite{DeFurio2022a}, $M_{\rm pri}\leq1.2 M_\odot$ in \cite{DeFurio2022b}, $M_{\rm total}\lesssim 1.5 M_\odot$ in \cite{calissendorff2022}, and $M_{\rm pri}\sim0.5-0.6 M_\odot$ in \cite{Biller2022}.
As suggested in \cite{Reggiani2011}, the mass ratios of the companions detected from our observations show consistency in the empirical mass ratio of M-type binary systems \citep[e.g. peak at $q=0.63-1$;][]{Fischer1992} rather than the initial mass function for a single M-type star \citep[peak at $q=0.25-0.4$;][]{bochanski2010}. 
The mass ratios between the primaries and the secondaries can be detailed in the forthcoming paper(s) after obtaining follow-up data, which are used in orbital fitting with the combination of the RV and the imaging information.
The direct spectroscopy fed by AO corrections such as Subaru/REACH \citep{Kotani2020} with a Subaru NIR wavefront sensing \citep{Lozi2022} or Keck/KPIC \citep{Mawet2016} is also useful to reveal the mass ratios.  
The astrometric information from {\it Gaia} may also be useful to better constrain the orbits \citep[e.g.][]{Feng2022}, with a caveat that incorporating {\it Gaia} astrometry of a high proper-motion star into the RV+direct-imaging orbital fitting might cause unreasonable fitting results \citep[][]{Biller2022}. 

All the resolved companions in our observations have projected separations smaller than 20~au while we did not detect any companions at larger separations within 4\arcsec, which is consistent with the peak of the separation distributions around M~dwarfs in \cite{Winters2019} rather than an M-type binary model in \cite{Susemiehl2022}. Although the mass and separation range between our observations and \cite{Susemiehl2022} is different, our results perhaps suggest unexplored populations at small separations around nearby M~dwarfs. 
Since the systems are all nearby ($\sim$10--30~pc) and the semi-major axes appear to be quite small (for example, LSPM J0825+6902, LSPM J1645+0444, and LSPM J2204+1505 all have companions whose projected separations 
are close to 2--3 au), they are good candidates for obtaining both individual brightnesses and individual dynamical masses for each stellar component. Such types of systems remain rare, but they are important for calibrating evolutionary and isochronal models in the low-mass stellar regime \citep[e.g.][]{calissendorff2022}.

\subsection{Non-detection Targets}

\begin{figure*}
\begin{tabular}{cc}
\begin{minipage}{0.5\hsize}
    \centering
    \includegraphics[width=0.85\textwidth]{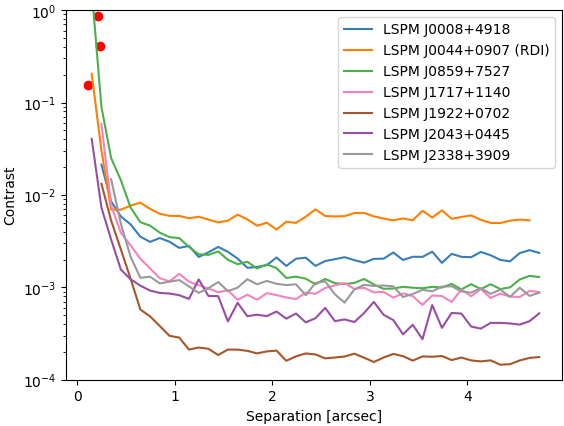}
\end{minipage}
\begin{minipage}{0.5\hsize}
    \centering
    \includegraphics[width=0.85\textwidth]{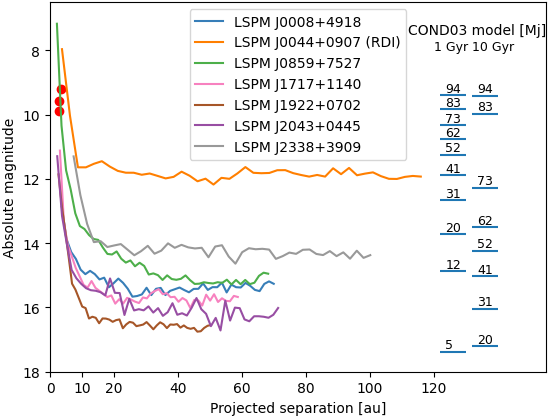}
\end{minipage}
\end{tabular}
\caption{Left: 5$\sigma$ contrast limit of the post-processed $L^\prime$-band data in our observations. We utilized RDI for the LSPM~J0044+0907 data set and ADI for the other data sets. We also present the detected companions from the NIRC2 $L^\prime$-band observations (LSPM~J0825+6902~B, LSPM~J1645+0444~B, an LSPM~J2204+1505~B) indicated by red circles. 
Right: Converted mass limit from the contrast limit overlaid with the detected companions as shown in the left panel and the COND03 evolutionary model assuming 1~Gyr and 10~Gyr. 
}
    \label{fig: contrast limit}
\end{figure*}

We did not detect any companion candidates within 4$^{\prime\prime}$ around LSPM~J0008+4918, LSPM~J0044+0907, LSPM~J0859+7257, LSPM~J1717+1140, LSPM~J2043+0445, LSPM~J1922+0702, and LSPM~J2338+3909 even though we conducted deep $L^\prime$-band ADI and RDI observations. 
Then, the colors of companion candidates around LSPM~J1922+0702 favor that they are field stars with the distances larger than the target star (see Figure \ref{fig: color}).
We also did not resolve any bright companions around LSPM~J0336+1350 and LSPM~J1534+1800 in the shallow Br$\gamma$ and $K^\prime$-band observations.  
Figure \ref{fig: contrast limit} shows the detection limits from the deep $L^\prime$-band observations including the LSPM~J1922+0702 data set where the two faint candidates were identified as background stars. 
We also used a cross-correlation function in the {\tt pyklip} modules to calculate the contrast limits after smoothing with a Gaussian ($\sigma$=3~pix). In these calculations, we performed injection tests to correct for self-subtraction induced by the ADI/RDI post-processing. 
Compared with the COND03 evolutionary model \citep{Baraffe2003} assuming an age range of 1-10~Gyr, our observations could set constraints on potential substellar-mass companions ($\sim10-75\ M_{\rm Jup}$ at 1-10~Gyr) at 10~au or further. 
Considering the fact that our targets have the RV trends or large RUWE values suggesting unseen companions, our null detections suggest the presence of closer-in or smaller-mass companions. Particularly LSPM~J0044+0907, LSPM~J1717+1140, and LSPM~J2043+0445 have the robust RV trends (13, 11, and 9$\sigma$ respectively) and large RUWE values (1.61, 5.7, and 10.2 respectively). 
Continuous monitoring of RV and future {\it Gaia} data releases of these targets will reveal the nature of these perturbations, which may also promote deeper imaging with JWST.
LSPM~J1534+1800 shows the large RV trend (6.6$\sigma$) and RUWE value (2.6) suggesting an unseen companion: Deep high-contrast imaging observations (which have not been performed for this system) as well as further RV monitoring will be able to investigate the potential companion.

\section{Summary} \label{sec: Summary}

The Subaru strategic program employing Subaru/IRD is performing a large monitoring campaign of RV measurements targeting $>120$ nearby M~dwarfs since 2019, where signatures of massive companions such as RV trends are deprioritized. However, this information remains useful for studying the stellar multiplicity of M~dwarfs. 
To search for outer companions around the IRD-SSP targets, we have conducted synergetic explorations with Keck/NIRC2 pyramid wavefront sensing, enabling high-contrast AO performance on nearby M~dwarfs.
This survey is the first high angular resolution/contrast imaging explorations for companions around cool M~dwarfs using NIR wavefront sensing.
We selected the deprioritized IRD-SSP targets that have signatures of long-term RV trends and/or large RUWE values from the {\it Gaia} catalog suggesting unresolved systems.

We observed 14 nearby M~dwarfs with Keck/NIRC2 and eventually detected companion candidates around five M~dwarfs and confirmed one known companion.
The two-epoch astrometry of the companion candidates around LSPM~J1002+1459, LSPM~J2204+1505, and LSPM~J1992+0702 revealed that LSPM~J1002+1459~B and LSPM~J2204+1505~B are real companions while the two faint candidates around LSPM~J1992+0702 are background stars.  
For the rest of the candidates, we calculated the FAPs using the 2MASS catalog, which suggests that they are likely companions with $>4\sigma$ confidence. 
Including the known companions resolved by the IRD-FIM camera, we detected seven companions between $\sim2-20$~au in total (four new companions between $\sim2-10$~au).
The $K-L^\prime$ colors of the companions around LSPM~J0825+6902, LSPM~J1645+0444, and LSPM~J2204+1505 are consistent with low-mass field stars.
LSPM~J0825+6902~B and LSPM~J2204+1505~B are located at the boundary between late-M and early-L spectral types; they are good targets for follow-up characterizations including spectrophotometric observations and orbit analysis. 

Even with the deep $L^\prime$-band ADI/RDI explorations for LSPM~J0008+4918, LSPM~J0044+0907, LSPM~J0859+7257, LSPM~J1717+1140, LSPM~J2043+0445, and LSPM~J2338+3909, we did not detect any additional companion candidates within 4$^{\prime\prime}$.
We then calculated 5$\sigma$ detection limits compared with the COND03 evolutionary model, which could set constraints on potential substellar-mass companions at 10~au or further. Given that our targets have suggestions of unseen companions, null detection of Keck/NIRC2 observations favors inner-orbit and smaller-mass companions; continuous monitoring of RV and/or deeper imaging with JWST will reveal the nature of these perturbations.


\vspace{1in}

The authors would like to thank the anonymous referee for their constructive comments and suggestions to improve the quality of the paper.
We are grateful to Tomas Stolker who helped arrange the internal settings of {\tt species}.
T.U. is supported by Grant-in-Aid for Japan Society for the Promotion of Science (JSPS) Fellows and JSPS KAKENHI Grant No. JP21J01220. 
M.T. is supported by JSPS KAKENHI grant Nos. 18H05442, 15H02063, and 22000005. T.H. is supported by JSPS KAKENHI Grant Nos. 19K14783 and 21H00035. E.K. is supported by JSPS KAKENHI grant No. 18H05438.
This research is partially supported by NASA ROSES XRP award 80NSSC19K0294, JSPS KAKENHI Grant No. JP18H05439, and JST CREST Grant No. JPMJCR1761. 
Some of the research described in this publication was carried out at the Jet Propulsion Laboratory, California Institute of Technology, under a contract with the National Aeronautics and Space Administration. 
\par
The part of data presented herein were obtained at the W. M. Keck Observatory, which is operated as a scientific partnership among the California Institute of Technology, the University of California and the National Aeronautics and Space Administration. The Observatory was made possible by the generous financial support of the W. M. Keck Foundation.
Part of this research is based on data collected at the Subaru Telescope, which is operated by the National Astronomical Observatory of Japan. 
The authors wish to recognize and acknowledge the very significant cultural role and reverence that the summit of Maunakea has always had within the indigenous Hawaiian community.  We are most fortunate to have the opportunity to conduct observations from this mountain.
\par
This work has made use of data from the European Space Agency (ESA) mission
{\it Gaia} (\url{https://www.cosmos.esa.int/gaia}), processed by the {\it Gaia}
Data Processing and Analysis Consortium (DPAC,
\url{https://www.cosmos.esa.int/web/gaia/dpac/consortium}). Funding for the DPAC
has been provided by national institutions, in particular the institutions
participating in the {\it Gaia} Multilateral Agreement.
This research has made use of NASA's Astrophysics Data System Bibliographic Services.
This research has made use of the SIMBAD database, operated at CDS, Strasbourg, France.
Our modeling of the distortion of IRD's FIM camera used M5 images that benefited from observations made with the NASA/ESA Hubble Space Telescope, and obtained from the Hubble Legacy Archive, which is a collaboration between the
Space Telescope Science Institute, the Space Telescope
European Coordinating Facility (ST-ECF/ESA) and the
Canadian Astronomy Data Centre (CADC/NRC/CSA).

\software{\texttt{astropy}: \citet{astropy:2013,astropy:2018}}


\bibliography{library}                                    
\end{document}